\documentclass[12pt]{article}

\usepackage{amssymb,amsmath, psfrag, epsfig}
\usepackage{caption}
\usepackage[section]{placeins}
\def\o{\over}

\def\p{\partial}
\def\ov{\overline}
\def\b{\noindent}  

\def\nh{\noindent\hangindent=1 true cm \hangafter = 1}

\def\nh{\noindent\hangindent=1 true cm \hangafter = 1}

\def\B {\begin{eqnarray*}}

\newcommand{\bel}[1]{\begin{equation}\label{#1}}

\newcommand{\be}{\begin{equation}}
\newcommand{\qe}{\end{equation}}
\newcommand{\eeq}{\end{equation}}
\newcommand{\baS}{\begin{eqnarray}}
\newcommand{\ba}{\begin{eqnarray}}
\newcommand{\ea}{\end{eqnarray}}

\def\EN{\end{eqnarray*}}  

\begin{document}
\title{The space-clamped Hodgkin-Huxley system with random synaptic input: inhibition of spiking by weak noise and analysis with moment equations} 
\author{ Henry C. Tuckwell$^{1,2}$ and Susanne Ditlevsen$^3$\\
tuckwell@adelaide.edu.au\\
$^1$ School of Electrical and Electronic Engineering, University of
Adelaide\\ SA 5005, Australia \\
\\
$^2$School of Mathematical Sciences, Monash University\\
Clayton, Victoria 3168, Australia\\
\  \\
$^3$Department of Mathematical Sciences, University of Copenhagen\\
DK 2100 Copenhagen $\phi$, Denmark\\
\  \\
\ \\   
{\it Email addresses}\\
tuckwell@adelaide.edu.au;  susanne@math.ku.dk\\\\ 
\   \\}

\maketitle

\begin{abstract}  
We consider a classical space-clamped Hodgkin-Huxley model neuron stimulated by synaptic excitation and inhibition with conductances represented by Ornstein-Uhlenbeck processes. Using numerical solutions of the  stochastic model system obtained by an
Euler method, it is found that with excitation only there is a critical
value of the steady state excitatory conductance for repetitive spiking
without noise 
and for values of the conductance near the critical value small noise
has a powerfully inhibitory effect. For a given level of inhibition there is also a critical value of the steady state excitatory conductance for repetitive firing and it is demonstrated that noise either in the excitatory or inhibitory processes or both can powerfully inhibit spiking. Furthermore, near the critical value, inverse stochastic resonance
was observed when noise was present only in the inhibitory input process. 

The system of 27 coupled deterministic differential equations for the approximate first and second order moments of the 6-dimensional model is derived. The moment differential equations are solved using Runge-Kutta methods and the solutions are compared with the results obtained by simulation for various sets of parameters including some with
conductances obtained by experiment on pyramidal cells of rat prefrontal cortex. The mean and variance obtained
from simulation are in good agreement when there is spiking induced by strong stimulation and relatively small noise or when the voltage is
fluctuating at subthreshold levels. In the occasional spike mode sometimes exhibited by spinal motoneurons and cortical pyramidal cells the assunptions underlying
the moment equation approach are not satisfied.

\end{abstract}
\newpage

\tableofcontents

\rule{60mm}{1.5pt}

\section{Introduction}
The stochastic nature of neuronal discharges was first reported in the 1930s, through experiments which found variability in the responses of  frog myelinated axon to shocks of the same intensity and duration, notably by Monnier and Jasper (1932), Blair and Erlanger (1932) and Pecher (1936, 1937). Concerning the last mentioned author, Verveen has documented an interesting account
of his career, which started in Belgium and ended mysteriously at the age of 28 in the United States after researching radioactive substances
whose nature was considered a military secret. (See www.verveen.eu which also has links to 
Pecher's articles (in French)). Some notable later contributions, including review, on this aspect of
stochasticity in neurons are Verveen (1960), Lecar and Nossal (1971) 
and Clay (2005).

Since these early discoveries,  there have been a very large number of experimentalist studies which revealed the stochastic nature of nearly all neuronal activity, the latter term referring mainly to action potential generation. This embraces single neurons and their component parts and populations of neurons and glia. Some of the pioneering works were on interspike interval variability in muscle spindles 
(Brink et al., 1946), cells in the auditory  system (Gerstein and Kiang, 
1960) and those in the visual cortex (Burns and Webb, 1976). 

Mathematical modeling of the complex dynamical processes underlying such activity has since flourished (Bachar et al., 2013).  Much of the theoretical work has focused on channel noise at both the single channel level (Colquhoun and Hawkes, 1981) and at the patch level modeled by a diffusion limit for a birth and death process (Tuckwell, 1987). The role of such noise in   
the generation of action potentials, for example by altering the firing threshold,  has been explored in (for example) White et al. (2000),  
 Austin (2008) and Li et al. (2010). 

Linear models with synaptic noise have been studied since the 1960s and there have been more recent
analyses of them such as those of Hillenbrand (2002), Ditlevsen (2007) and Berg and Ditlevsen (2013).  Most nonlinear models of neuronal activity are based on
Hodgkin- Huxley (1952) type systems and include spatial models (Horikawa, 1991; Sauer and Stannat, 2016).
 Some recent studies include those of Wenning et al. (2005), who employed a model similar to that explored in the present article,  and Finke et al. (2008), who included both additive noise and channel noise in the form of Ornstein-Uhlenbeck
processes  (OUPs) for some activation variables, and Yi et al. (2015),
who studied noise effects on spike threshold in a two-dimensional
Hodgkin-Huxley (HH) type model 
with synaptic noise represented by an OUP.

The focus of the present article is on the classical HH system with
synaptic noise and its analysis by the moment method, in a similar
vein to  
that for the same system with additive white noise (Rodriguez and Tuckwell, 2000; Tuckwell and Jost, 2009).
With weak additive white noise or with conductance-based noise as in the present model (Tuckwell et al., 2009) or with colored noise (Guo, 2011), and mean input currents near the threshold for repetitive firing, the firing rate undergoes a minimum (inverse stochastic resonance) as the noise level increases from zero. We examine the responses of the system to synaptic input, either excitatory alone, or with inhibition near the threshold for repetitive spiking and compare solutions obtained by the moment method with simulation.   The moment method and extensions of it  has been employed in several recent studies of neuronal networks (Deco and Marti, 2007; Mareiros et al., 2009; Hasegawa, 2009; Deco et al., 2013; 
Franovi\'c et al., 2013; Hasegawa, 2015) and genetic networks
(Sokolowski and Tka$\check{\rm c}$ik, 2015).

\section{The HH equations with random synaptic currents}

The standard HH system consists of four ordinary differential equations for the electrical
potential $V(t)$, the potassium activation variable $n(t)$, the sodium activation variable $m(t)$ and the 
sodium inactivation variable $h(t)$. The latter three variables take values in $[0,1]$ and their differential
equations involve the coefficients $\alpha_n, \beta_n, \alpha_m, \beta_m, \alpha_h, \beta_h$ which depend on $V$. In a previous
article
(Tuckwell and Jost, 2009), the HH system with an
additive current including deterministic and random components was
analyzed.  
Here we use similar techniques for the HH system with random
synaptic excitation and inhibition.  See also Tuckwell and Jost (2012) and Bashkirtsiva et al. (2015) for further analysis of the effects
of noise in the HH model with additive noise.

Random synaptic inputs can be included in several ways (e.g.,
Rodriguez and Tuckwell, 1998), but we restrict our attention to cases
where there are no 
discontinuities in $V$ but rather the synaptic currents can be
represented with diffusion processes, which makes the mathematical
formalism somewhat less complicated. The resultant system 
consists of 6 coupled stochastic differential equations (SDEs) where the
stochasticity is only explicit in the 5th and 6th equations.

The  form of the systems under consideration here is that of an
 $n$-dimensional temporally homogeneous diffusion process ${\bf X}= \{{\bf X}(t), t \ge 0\}$
where the  $j$-th component satisfies the stochastic differential equation 
\be dX_j(t) = f_j({\bf X}(t))dt + \sum_{k=1}^{m} g_{jk}({\bf X}(t)) dW_k(t), \qe
where the $W_k$s are independent standard (zero mean, variance $t$ at time $t$) Wiener processes 
and it is assumed that conditions for
existence and uniqueness of solutions are met (Gihman and Skorohod, 1972). 

\subsection{Description of model}

\b For the present model 
 we let the first four variables be $X_1=V$, $X_2=n$, $X_3=m$ and $X_4=h$ and 
${\bf X}$ be the vector of all 6 components. The first equation of a general HH system with synaptic inputs can be written
\be  dX_1 =\frac{1}{C}  \bigg(  H({\bf X}) + I_{syn} \bigg)dt  \qe
 where $I_{syn}$ represents the synaptic input currents and 
we have defined 
 the sum of the standard HH currents as 
\be H({\bf X}) = \ov{g}_KX_2^4(V_K-X_1)  +  \ov{g}_{Na}X_3^3X_4(V_{Na}-X_1) + g_L(V_L-X_1),  \qe 
where  $\overline{g}_K$, $\overline{g}_{Na}$ and $g_L$ are the maximal (constant) potassium, 
           sodium and leak conductances per unit area with corresponding equilibrium potentials $V_K$, $V_{Na}$,
           and $V_L$, respectively.
The 2nd, 3rd and 4th equations take the standard HH forms, 
\begin{eqnarray} 
dX_2\,= \, f_2({\bf X})dt &=& \big[\alpha_n(X_1)(1-X_2) - \beta_n(X_1)X_2\big]dt \\[2mm] 
dX_3\,=\, f_3({\bf X})dt &=& \big[\alpha_m(X_1)(1-X_3) - \beta_m(X_1)X_3\big]dt \\[2mm] 
dX_4\,=\, f_4({\bf X})dt &=& \big[\alpha_h(X_1)(1-X_4) -  \beta_h(X_1)X_4\big]dt 
\end{eqnarray}
In the following we denote the derivatives of $\alpha_n$ etc. by $\alpha_n'$ etc.

A representation of certain synaptic inputs has been
successfully used in models for the spontaneous random spiking of cat and rat neocortical pyramidal neurons (Destexhe et al., 2001; Fellous et al., 2003) in which the excitatory
and inhibitory inputs constituting the ongoing background input are presumed to be composed mainly
of small and frequent  glutamatergic and GABA-ergic postsynaptic
currents which could be separately identified.  A similar model was 
employed for an HH neuron with conductance-driven input by Tuckwell
et al. (2009) in a demonstration of the robustness of the phenomenon 
of inverse stochastic resonance. In the Destexhe et al. (2001) model
there is, in addition to the usual three HH currents, an M-type potassium
current which, as in Tuckwell et al. (2009), is not included here. In actual neocortical pyramidal cells there are
several more component currents than in the Destexhe et al. (2001)
model - see for example Yu et al. (2008) for a partial list.
The overall synaptic current is then written by Destexhe et al. (2001) as
\be I_{syn}= \frac{1}{A} [g_e(t)(V_E-V) + g_i(t)(V_I-V)], \qe
where  $V_E, V_I$ are equilibrium potentials for excitatory and inhibitory 
synaptic input, the synaptic excitatory and inhibitory conductances 
at time $t$ are $g_e(t)$ and $g_i(t)$, respectively, and $A$ is the
 ``total membrane area''. 
The synaptic 
conductances are ascribed the stochastic differential equations
\be dg_{e}(t)=  -\frac{1}{\tau_{e}}(g_e(t)-\bar{g}_e)dt + \sigma_edW_e(t) \qe
\be dg_{i}(t)= -\frac{1}{\tau_{i}}(g_i(t)-\bar{g}_i)dt + \sigma_idW_i(t) \qe
where $\tau_e, \tau_i$ are time constants, $\bar{g}_e, \bar{g}_i$ are equilibrium
values, 
 $W_e, W_i$ are 
corresponding (assumed independent) standard Wiener processes and $\sigma_e, \sigma_i$ are noise amplitudes.  Thus the processes
$g_e$ and $g_i$ are of the Ornstein-Uhlenbeck type. 

In this model, the first of the 6 SDEs is
simply
\be   dX_1=f_1({\bf X})dt  \qe
with
 \begin{eqnarray}
 f_1({\bf X}) &= & \frac{1}{C}  \bigg(H({\bf X}) + \frac{X_5}{A}(V_E-X_1) + \frac{X_6}{A}(V_I-X_1)   \bigg) , \notag  \\
 \end{eqnarray} 
where $X_5=g_e$ and $X_6=g_i$. 

For $X_5$ and $X_6$ we have 
\be dX_5 =f_5(X_5)dt + g_{51}dW_1 \qe
and
\be dX_6 =f_6(X_6)dt + g_{62} dW_2  \qe
where we identify $W_e = W_1$ and $W_i=W_2$, and
where
\be f_5(X_5)= -\frac{1}{\tau_{e}}(X_5-\bar{g}_e) \qe
\be f_6(X_6)= -\frac{1}{\tau_{i}}(X_6-\bar{g}_i). \qe
Further  $g_{51}=\sigma_e$ and $g_{62}= \sigma_i$ and these are
the only two non-zero $g_{jk}$s in this model so $m=2$ in Equ. (1). 
There are 27 distinct first and second order moments for this model and differential equations for their approximations are obtained in Section 3.

 \section{Simulation results}

\subsection{Simulations and the inhibitory effect of noise}
Simulated solutions of the system of 6 coupled SDEs  
equations defining the above HH model neuron with random 
synaptic inputs are obtained with a simple Euler method in which
discretization is applied with a time step of $\Delta t$.  Unless stated
otherwise, the value $A=1$ is employed in the calculations described
below. 

\subsubsection{Excitation: critical value and choice of $\Delta t$}
With no inhibition and no noise so that $\sigma_e=0$,
there is a critical value of $\bar{g}_e$ for repetitive firing.
When $\bar{g}_e=0.1$ repetitive (presumed continuing
indefinitely) firing does not occur for any value of $\Delta t$
tested.   Figure 1A  shows the voltage response without noise and
with $\Delta t=0.002$. There is only one spike at about $t \approx 5$. 
With very small noise  $\sigma_e=0.0005$, the results are shown in Figure 1B for 10 trials,
the response being almost identical in each trial. When the noise
amplitude is greater at $\sigma_e=0.005$ as in Figure 1C, second spikes
may emerge as is the case in 4 of the ten trials depicted.
Figure 1D shows 4 voltage trajectories   for a still greater noise level
$\sigma_e=0.05$. Here there are in all cases (many not shown) an
apparently unceasing sequence of randomly occurring spikes, despite the fact that the value of $\bar{g}_e$ is less than the critical value for
repetitive firing without noise.

\begin{figure}[!h]
\begin{center}
\centerline\leavevmode\epsfig{file=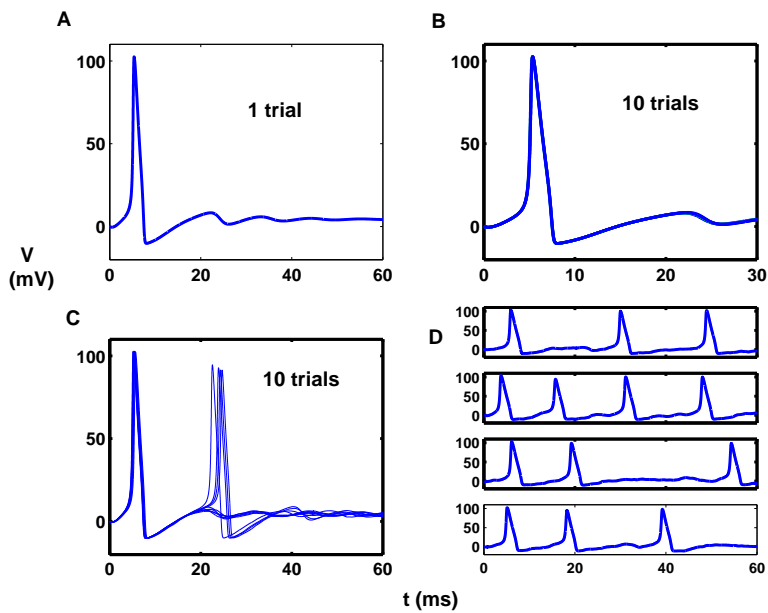,width=4.0in}
\end{center}
\caption{Simulation results with excitation only and various
noise levels. In all cases
steady state excitation  $\bar{g}_e=0.10$ below the
critical value for repetitive firing and a time step of $\Delta t=0.002$.
A. No noise (1 trial) giving one spike. B. Ten trials with noise
$\sigma_e = 0.0005$.  One spike only, being practically the same
on each trial.  C. Ten trials with $\sigma_e=0.005$.  The noise is
sufficient to give rise to a second spike in 4 of the ten trials.
 D.  4 individual trials with $\sigma_e=0.05$. In all cases examined
multiple sustained spiking at random times occurs.} 
\label{fig:wedge}
\end{figure}

Using the previously found (Tuckwell et al., 2009) critical value $\bar{g}_e=0.112$ as a guide, spike trains were examined for 
values of $\bar{g}_e$ close to that value for various values of $\Delta t$.
Results will be reported only for the two values $\Delta t=0.015$ and
$\Delta t=0.002$.  For $\bar{g}_e=0.111$, there were 4 and 6 spikes
respectively for the smaller and larger time steps, and for the slightly
larger value $\bar{g}_e=0.1115$ there were 4 spikes of declining
amplitude for the smaller time step whereas for the larger time step
there were 13 spikes in 240 ms whose amplitudes finally remained
constant - see Figure 2A. With $\bar{g}_e=0.112$ there were 7
spikes of declining amplitude for the smaller time step and an
apparently repetitive train for the larger time step.

Finally, with $\bar{g}_e=0.1125$ just above the previously
determined critical value
an apparently stable repetitive spike train was  obtained with
both the larger and smaller time step. This is seen in Figure 2 and
suggests that the critical value is very close to 0.1125.
In general it was observed that the larger time step sped up the
spiking and tended to make it more stable. However, 
the above results indicated that it is preferable to use the
smaller time step despite its leading to significantly greater computation times. 

\begin{figure}[!h]
\begin{center}
\centerline\leavevmode\epsfig{file=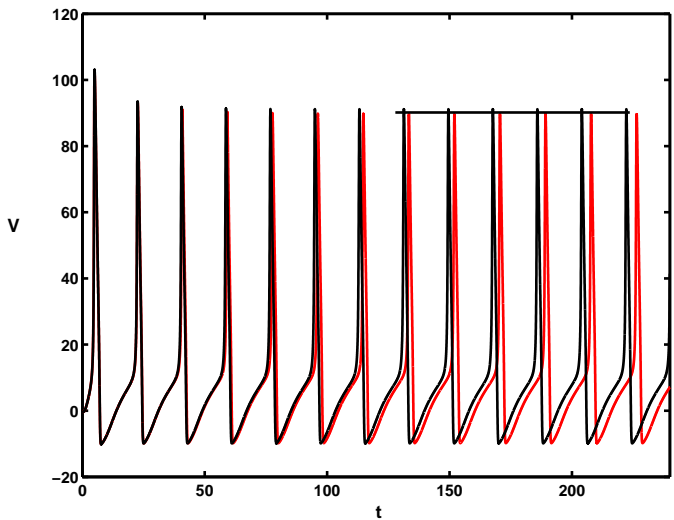,width=3.5in}
\end{center}
\caption{Spike trains for $\bar{g}_e=0.1125$ just above the
critical value for repetitive spiking, showing the effects of 
smaller (red) and larger (black) time steps of 0.002 and 0.015
respectively.  The larger time step leads to a higher frequency. Horizontal segment indicates stable spike amplitudes
for both time steps.} 
\label{fig:wedge}
\end{figure}

\subsubsection{Inhibition by noise with excitation only}
With excitation only at  $\bar{g}_e=0.1125$, so that without 
noise there is repetitive periodic firing, noise of a small
amplitude can lead to a greatly reduced number of spikes.
This is illustrated in Figure 3 where the three columns
show 4 trials, of length 100 ms, for each of three values of $\sigma_e$, being
0.0025, 0.01 and 0.025, increasing from left to right.
With no noise there are 6 spikes (see Figure 2). 
The smallest of the values of $\sigma_e$ leads to the
greatest reduction in average spike numbers, to 3.0, whereas
the largest value of $\sigma_e$ has an average spike number of
5.75. 

\begin{figure}[!h]
\begin{center}
\centerline\leavevmode\epsfig{file=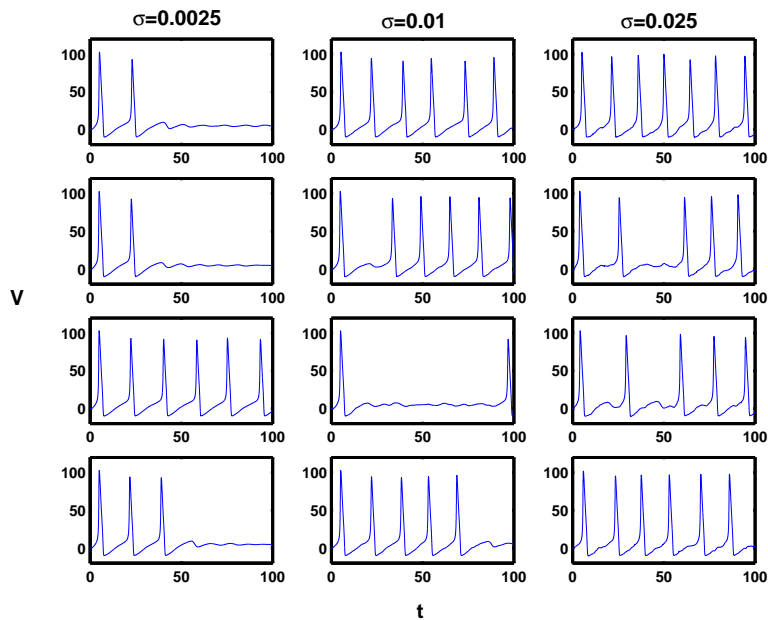,width=4in}
\end{center}
\caption{Spike trains for $\bar{g}_e=0.1125$ just above the
critical value for repetitive spiking, showing the effects of small noise 
of three different magnitudes, increasing from left to right, in 4 trials.} 
\label{fig:wedge}
\end{figure}

\subsubsection{Inhibition by noise with excitation and inhibition} 
In previous investigations of the inhibitory effects of noise
on repetitive firing induced by synaptic input (Tuckwell et al., 2009), only excitatory
inputs have been considered. Here we briefly consider a few cases
in which the synaptic input is both excitatory and inhibitory and
there is noise in either the excitatory or inhibitory component or
both. 
 First it was required to find a combination of excitation and inhibition
which would lead to repetitive spiking in the deterministic case.
Without a formal proof, it seems that for any level of inhibition  $\bar{g}_i$
there can always be found a level of excitation  $\bar{g}_e$
to give repetitive spiking. With  $\bar{g}_i=0.1125$, close to the 
critical value of  $\bar{g}_e$ for excitation only, spike trains were
examined with a time step of 0.015 for various values of  $\bar{g}_e$.
The elicited trains for  $\bar{g}_e=0.1725$ and  $\bar{g}_e=0.1775$
did not exhibit repetitive spiking as shown in the first two panels
in the top row of Figure 4. When  $\bar{g}_e$ was increased slightly
to 0.1790, as in the right hand panel of the first row, repetitive
spiking was sustained, indicating that this level of excitation
was critical for the given level of inhibition. 
 In the remaining three rows of Figure 4 are shown the resulting spike trains with noise 
for three trials in which the values of  $\bar{g}_e$ and  $\bar{g}_i$
are those which gave repetitive spiking without noise. 
In the second row there is noise of a small magnitude $\sigma_e=0.0025$ in the excitation only, whereas in the third row
noise of magnitude   $\sigma_i=0.0025$ is present in the
inhibitory input only. Finally, in the 4th row there is noise
in both the excitatory and inhibitory input processes with $\sigma_e=\sigma_i=0.00125$  such that the
sum of the amplitudes is the same as in rows two and three.
The percentage reductions in average spike numbers,
for these (small) sample sizes, are 78, 67 and 64
for noise in excitation, inhibition and both, respectively.
This preliminary investigation indicates that inhibition of
repetitive spiking by noise is as strong when there is
inhibition present as when there is excitation only
 and occurs regardless of whether
the noise arises in excitatory, inhibitory or both input
processes.

\begin{figure}[!h]
\begin{center}
\centerline\leavevmode\epsfig{file=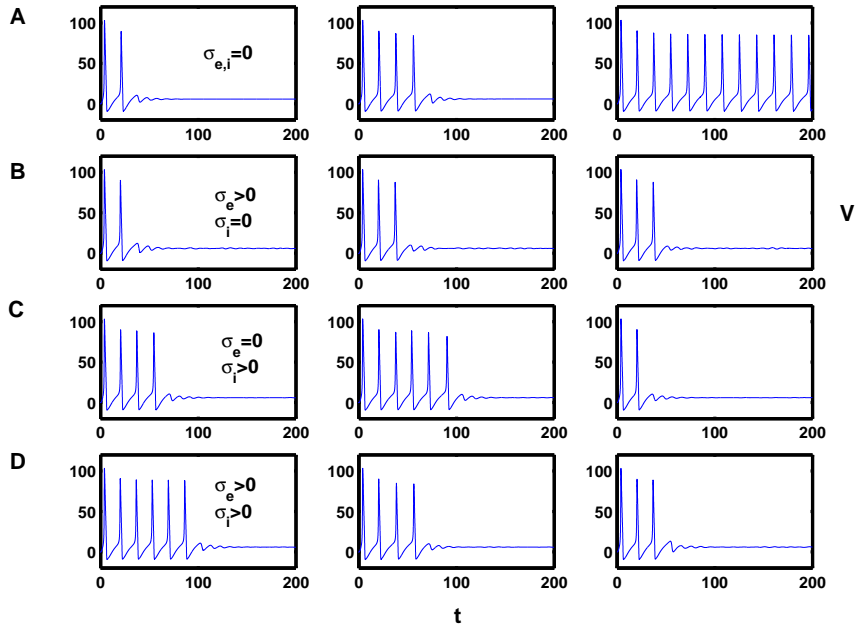,width=4.5in}
\end{center}
\caption{A. In the top three records there is no noise but
both excitatory and inhibitory inputs. In the left-most panel, with
$\bar{g}_e=0.1725$ and $\bar{g}_i=0.1125$,
 and in the middle panel with 
$\bar{g}_e=0.1775$ and $\bar{g}_i=0.1125$, repetitive firing is
not established. When, as in the right-hand panel,  the larger value $\bar{g}_e=0.1790$ is
employed with the same value of $\bar{g}_i$, repetitive firing
occurs. B.  In the results of the second row the values of 
$\bar{g}_e$ and  $\bar{g}_i$ are as in the right-hand panel
of the first row, leading to repetitive spiking in the absence of noise, 
but that now a small noise $\sigma_e=0.0025$ is added only to the excitatory component, giving rise to a large degree of inhibition.
C. As in B but that now noise with amplitude $\sigma_i=0.0025$
is added to only the inhibitory component, which also inhibits the
spiking. D. As in B and C except that the noise is spread equallly
amongst excitatory and inhibitory inputs with $\sigma_e=0.00125$
and $\sigma_i=0.00125$. Significant inhibition of spiking is
observed in this case also.} 
\label{fig:wedge}
\end{figure}

Further results were obtained with the levels of excitation and inhibition 
$\bar{g}_e$ and $\bar{g}_i$
which led to repetitive spiking as in Figure 4A. 50 trials of length 100 ms
were performed for 30 values of the inhibitory  noise parameter $\sigma_i$, from 0 to 0.1, with no noise in the excitatory process.
The mean number of spikes versus noise level is shown in Figure 5,
black circles, and it is seen that the spike rate undergoes a minimum
around values of $\sigma_i$ just less than 0.02. Away from the critical value for repetitive spiking,  when the level of the
excitatory input was increased to $\bar{g}_e=0.2106$ but with the same value of  $\bar{g}_i$, the firing rate undergoes a much weaker minimum
as $\sigma_i$ is increased from 0 to 0.1. These results parallel those
obtained previously for increasing the excitatory noise level (Tuckwell et al., 2009). Somewhat surprisingly, therefore, the phenomenon of inverse stochastic resonance occurs with increasing level of inhibitory noise alone.

\begin{figure}[!h]
\begin{center}
\centerline\leavevmode\epsfig{file=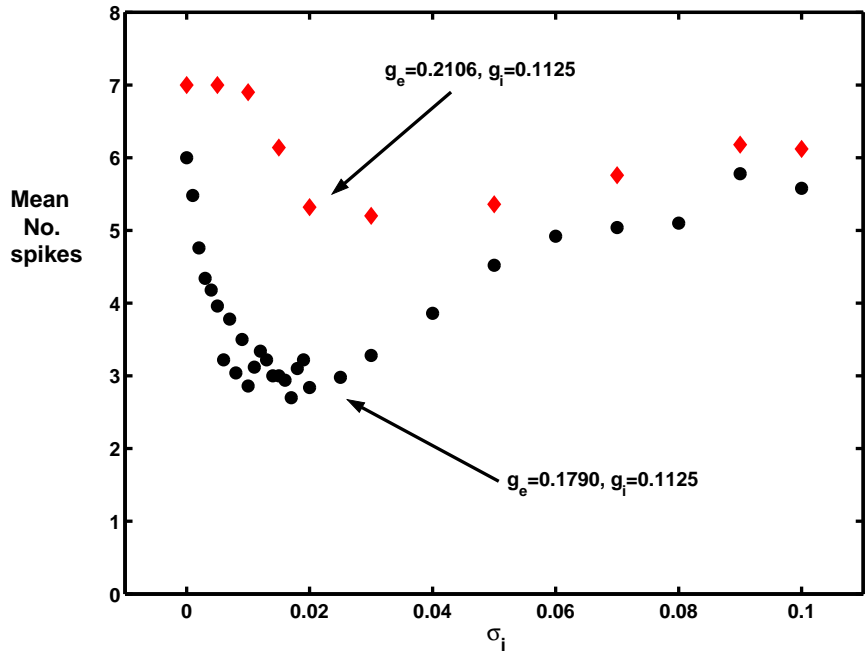,width=4.5in}
\end{center}
\caption{The mean number of spikes per trial is plotted against
the inhibitory noise parameter $\sigma_i$,  with no noise in the excitatory process, for values of $\bar{g}_e$ and $\bar{g}_i$ near a critical value for repetitive spiking (black circles) and at a level of excitation above the critical value (red diamonds). For the first set
a pronounced minimum occurs in the firing rate near $\sigma=0.02$,
providing evidence of inverse stochastic resonance with respect to
inhibitory noise alone.
} 
\label{fig:wedge}
\end{figure}

\section{Differential equations for the approximate first and second order moments}
We  will find deterministic differential equations 
satisfied by the approximations for the means, variances
and covariances of the components of the $n$-component vector-valued random process ${\bf X}$ in the above 
model, using the scheme
of Rodriguez and Tuckwell (1996).
Accordingly, for relatively small noise amplitudes
the exact means $\mu_i(t)= E[X_i(t)]$ and covariances
$K_{ij}(t) = E[(X_i(t)-\mu_i(t))(X_j(t)-\mu_j(t))]$  may sometimes be approximated by the functions  
$m_i(t), i=1,...,n$, and $C_{ij}(t), i,j = 1,...,n$, respectively,  which obey a system of
deterministic differential equations.   

The vector of means at time $t$ is denoted by ${\bf m}(t)$.  These quantities are found to satisfy the following
systems of ordinary differential equations
\be \frac{dm_i}{dt} = f_i({\bf m}) + \frac{1}{2} \sum_{l=1}^n\sum_{p=1}^n \bigg\{ \frac
{\p^2 f_i} {\p x_l \p x_p}\bigg\}_{({\bf m})} C_{lp} \qe
whereas the covariances are determined by

\begin{align}
  \label{eq:eight}
\begin{split}
\frac{dC_{ij} }{dt} = \ &  \sum_{k=1}^n \big\{g_{ik}g_{jk}\big\}_{({\bf m},t)}   + \sum_{l=1}^n \bigg\{   \frac{\p f_i}{\p x_l} \bigg\}_{({\bf m},t)} C_{lj}  +  
 \sum_{l=1}^n \bigg\{   \frac{\p f_j}{\p x_l} \bigg\}_{({\bf m},t)} C_{il} \\ 
 & +\frac{1}{2} \sum_{k=1}^n \sum_{l=1}^n\sum_{p=1}^n \bigg\{     g_{jk} \frac{\p^2 g_{ik}             } { \p x_l \p x_p              } 
            +       \frac{   \p g_{ik} } {  \p x_l   }  \frac{ \p g_{jk}    } { \p x_p }
            +  \frac{ \p g_{ik} } {\p x_p   } \frac {\p g_{jk}  }  { \p x_l  }  
            + g_{ik} \frac{\p^2 g_{jk}        } { \p x_l \p x_p      }      \bigg\}   _{({\bf m},t)}  C_{lp}.        
\end{split}
\end{align}

In the present model  this equation is simpler because the last line of Equ. (17) is absent, there being no triple  sum contribution
involving second derivatives or products of first order derivatives of the $g_{jk}$s because such terms are all constants, most of which are zero.  

In order to simplify the notation in the following equations we give some definitions. Let
$$H_1({\bf m})= \ov{g}_K  m_2^4 + \ov{g}_{Na} m_3^3m_4 + g_L +
 \frac{m_5}{A}  +  \frac{m_6}{A}$$
and let
\begin{align}
  \label{eq:eight}
  \begin{split}
H_2({\bf m}, y_1,y_2, y_3, y_4, y_5) =\ & \frac{4}{C}  \ov{g}_K  m_2^3 (V_K-m_1) y_1   + \frac{3}{C}  \ov{g}_{Na}m_3^2m_4(V_{Na}-m_1)y_2 \\
& + \frac{1}{C}  \ov{g}_{Na}m_3^3(V_{Na}-m_1)y_3 +  \frac{1}{AC} \big[(V_E-m_1)y_4 + (V_I-m_1)y_5\big]. 
  \end{split}
\end{align}

\subsection{The means}

Evaluating the required first and second order partial derivatives of $f_1,..,f_6$ we obtain the following differential equations
for the means $m_1,...,m_6$, of $V, n, m, h, g_i, g_e$, respectively.
For the voltage, 
 \begin{eqnarray*}
 \frac{dm_1}{dt}&=& \frac{1}{C}\bigg[H({\bf m})+ \frac{m_5}{A}(V_E-m_1) +  \frac{m_6}{A}(V_I-m_1)          \\ 
 &&  -4\ov{g}_Km_2^3 C_{12}  - 3\ov{g}_{Na}m_3^2m_4C_{13} -\ov{g}_{Na}m_3^3C_{14} -  \frac{C_{15}}{A}  -\frac{C_{16}}{A} \\
      && + 6 \ov{g}_Km_2^2(V_K-m_1)C_{22} + 3\ov{g}_{Na}m_3m_4(V_{Na}-m_1)C_{33} \\  
      && +  3\ov{g}_{Na}m_3^2(V_{Na}-m_1)C_{34} \bigg].  
           \end{eqnarray*} 
For the auxiliary variables
\begin{eqnarray*}
 \frac{dm_2}{dt}&=& \alpha_n(m_1)(1-m_2) - \beta_n(m_1)m_2
+ \frac{1}{2}  \big(\alpha_n''(m_1)(1-m_2)   - \beta_n''(m_1)m_2\big) C_{11}  \\  
      && -\big(\alpha_n'(m_1)+ \beta_n'(m_1)\big) C_{12}\\
 \frac{dm_3}{dt}&=& \alpha_m(m_1)(1-m_3) - \beta_m(m_1)m_3
      +  \frac{1}{2}  \big(\alpha_m''(m_1)(1-m_3)   - \beta_m''(m_1)m_3\big) C_{11}  \\  
      && -\big(\alpha_m'(m_1)+ \beta_m'(m_1)\big) C_{13} \\
 \frac{dm_4}{dt}&=& \alpha_h(m_1)(1-m_4) - \beta_h(m_1)m_4
 +  \frac{1}{2} \big(\alpha_h''(m_1)(1-m_4)   - \beta_h''(m_1)m_4\big) C_{11}  \\  
      && -  \big(\alpha_h'(m_1)+ \beta_h'(m_1)\big) C_{14}.  
\end{eqnarray*} 
For the means of the synaptic conductances we have      
          $$
 \frac{dm_5}{dt}=-\frac{1}{\tau_e} \big(m_5  -\bar{g}_e\big), \hskip .5 in
   \frac{dm_6}{dt}=-\frac{1}{\tau_i} \big(m_6  -\bar{g}_i\big).  
           $$
Since these are means of OUPs, exact solutions are known, which in fact coincide with the approximations. Solutions are 
$$m_5(t)=m_5(0)e^{t/\tau_e} + \bar g_e (1-e^{t/\tau_e})$$
and likewise for $m_6(t)$.

    \subsection{The variances}

                   For the above model of an HH system with random synaptic
input,  the differential equations for the variances of $X_i(t)$, obtained by 
  substituting the appropriate derivatives and coefficients are as follows.
For the variance of $V$  we have
                          \begin{eqnarray*}
           \frac{dC_{11}}{dt}&=& -\frac{2}{C} H_1({\bf m}) C_{11}
            + \frac{8}{C}\ov{g}_Km_2^3(V_K-m_1)C_{12} + \frac{6}{C}   \ov{g}_{Na}m_3^2m_4(V_{Na}-m_1)C_{13}\\
            &&    + \frac{2}{C}  \ov{g}_{Na}m_3^3(V_{Na}-m_1)C_{14}  +
            \frac{2}{AC} \big[(V_E-m_1)C_{15} + (V_I-m_1)C_{16}\big]. 
              \end{eqnarray*} 
For the variances of $n$, $m$ and $h$ we find
               \begin{eqnarray*}
           \frac{dC_{22}}{dt}&=& 2  \bigg[ \big(  \alpha_n'(m_1)(1-m_2) - \beta_n'(m_1)m_2\big)C_{12}
          -   \big( \alpha_n(m_1) +  \beta_n(m_1) \big) C_{22}  \bigg] \\[2mm]
           \frac{dC_{33}}{dt}&=& 2\bigg[     \big( \alpha_m'(m_1)(1-m_3) - \beta_m'(m_1)m_3\big)    C_{13}
                -  \big(\alpha_m(m_1)+ \beta_m(m_1)\big) C_{33}\bigg]    \\[2mm]
           \frac{dC_{44}}{dt}&=& 2\bigg[     \big( \alpha_h'(m_1)(1-m_4) - \beta_h'(m_1)m_4\big)    C_{14}
         -  \big(\alpha_h(m_1)+ \beta_h(m_1)\big) C_{44} \bigg]. 
                     \end{eqnarray*} 
For the variances of the excitatory and inhibitory conductances we find 
   $$        \frac{dC_{55}}{dt}= \sigma_e^2 -  \frac{2}{\tau_e}C_{55}, \hskip .5in
           \frac{dC_{66}}{dt}= \sigma_i^2 -  \frac{2}{\tau_i}C_{66}.  $$
Since these variances are for OUPs we know the exact solutions
$$C_{55} (t)= \frac{\sigma_e^2 \tau_e }{2}(1-e^{2t/\tau_e})$$
and likewise for $C_{66}(t)$, coinciding with the approximations.

\subsection{Remaining covariances}
The remaining 15 covariances  are solutions of the following differential equations.   
               \begin{eqnarray*}
           \frac{dC_{12}}{dt}&=& \big[  \alpha_n'(m_1)(1-m_2) - \beta_n'(m_1)m_2\big]C_{11} - \big[\alpha_n(m_1)
            + \beta_n(m_1)\big] C_{12}\\
            &&-\frac{1}{C} H_1({\bf m})  C_{12}  +H_2({\bf m}, C_{22},
     C_{23}, C_{24}, C_{25}, C_{26}) \\[2mm]
%
           \frac{dC_{13}}{dt}&=& \big[  \alpha_m'(m_1)(1-m_3) - \beta_m'(m_1)m_3\big]C_{11}  -[\alpha_m(m_1)
            + \beta_m(m_1)] C_{13} \\
&& - \frac{1}{C}  H_1({\bf m}) C_{13}  +  H_2({\bf m}, C_{23},
     C_{33}, C_{34}, C_{35}, C_{36})  \\[2mm]
%
           \frac{dC_{14}}{dt}&=& [\alpha_h'(m_1)(1-m_4) - \beta_h'(m_1)m_4]C_{11} -   [\alpha_h(m_1)
            + \beta_h(m_1)] C_{14} \\
            &&- \frac{1}{C}  H_1({\bf m})C_{14} +  H_2({\bf m}, C_{24},
     C_{34}, C_{44}, C_{45}, C_{46}) \\[2mm]
%
           \frac{dC_{15}}{dt}&=& - \frac{1}{C} H_1({\bf m}) C_{15} 
 +  H_2({\bf m}, C_{25},
     C_{35}, C_{45}, C_{55}, C_{56}) - \frac{1}{\tau_e}C_{15}\\[2mm]
%
  \frac{dC_{16}}{dt}&=& - \frac{1}{C} H_1({\bf m})  C_{16} 
 H_2({\bf m}, C_{26},
     C_{36}, C_{46}, C_{56}, C_{66})- \frac{1}{\tau_i}C_{16}\\
           \frac{dC_{23}}{dt}&=& \big( \alpha_m'(m_1)(1-m_3) - \beta_m'(m_1)m_3\big)    C_{12}
                    + \big(\alpha_n'(m_1)(1-m_2) - \beta_n'(m_1)m_2\big) C_{13}\\
                    &&      -      \big( \alpha_m(m_1) + \beta_m(m_1) + \alpha_n(m_1) + \beta_n(m_1) \big) C_{23} \\   [2mm] 
%
           \frac{dC_{24}}{dt}&=& \big( \alpha_h'(m_1)(1-m_4) - \beta_h'(m_1)m_4\big)    C_{12}
+ \big(\alpha_n'(m_1)(1-m_2) - \beta_n'(m_1)m_2\big) C_{14}\\
                    && - \big(\alpha_n(m_1) +\beta_n(m_1) + \alpha_h(m_1)+ \beta_h(m_1) \big)  C_{24}\\[2mm]
%
                   \frac{dC_{25}}{dt}&=& \big(\alpha_n'(m_1)(1-m_2) - \beta_n'(m_1)m_2\big) C_{15}
 -      \big(\alpha_n(m_1) +\beta_n(m_1) \big)  C_{25}  - \frac{1}{\tau_e}C_{25}\\[2mm]
%
                   \frac{dC_{26}}{dt}&=& \big(\alpha_n'(m_1)(1-m_2) - \beta_n'(m_1)m_2\big) C_{16}
    -      \big(\alpha_n(m_1) +\beta_n(m_1) \big)  C_{26}  - \frac{1}{\tau_i}C_{26}\\[2mm]
%
 \frac{dC_{34}}{dt} &= & \big( \alpha_h'(m_1)(1-m_4) - \beta_h'(m_1)m_4\big)    C_{13}    + \big(\alpha_m'(m_1)(1-m_3) - \beta_m'(m_1)m_3\big) C_{14}\\
               &&    -      \big(\alpha_m(m_1) +\beta_m(m_1)
               +\alpha_h(m_1) + \beta_h(m_1) \big)    C_{34} \\[2mm]
\frac{dC_{35}}{dt}&=& \big( \alpha_m'(m_1)(1-m_3) - \beta_m'(m_1)m_3\big)    C_{15}     -      \big( \alpha_m(m_1) + \beta_m(m_1) \big) C_{35}  - \frac{1}{\tau_e}C_{35}\\[2mm]
 \frac{dC_{36}}{dt}&=& \big( \alpha_m'(m_1)(1-m_3) - \beta_m'(m_1)m_3\big)    C_{16}     -      \big( \alpha_m(m_1) + \beta_m(m_1) \big) C_{36}  - \frac{1}{\tau_i}C_{36} \\ [2mm]
\frac{dC_{45}}{dt}&=& \big( \alpha_h'(m_1)(1-m_4) - \beta_h'(m_1)m_4\big)    C_{15}
                 -      \big(\alpha_h(m_1)+ \beta_h(m_1) \big) C_{45}  - \frac{1}{\tau_e}C_{45} \\[2mm]
\frac{dC_{46}}{dt}&=& \big( \alpha_h'(m_1)(1-m_4) - \beta_h'(m_1)m_4\big)    C_{16}
              -      \big(\alpha_h(m_1)+ \beta_h(m_1) \big) C_{46}  - \frac{1}{\tau_i}C_{46} \\[2mm]
             \frac{dC_{56}}{dt}&=&   -\bigg( \frac{1}{\tau_e} +  \frac{1}{\tau_i}     
\bigg)  C_{56}
                 \end{eqnarray*} 
However, the correct solution of the last equation must be $C_{56}=0$  since the Wiener processes $W_e$ and $W_i$ are independent. 
Hence the terms in $H_2$ in  the equations for
$C_{15}$ and $C_{16}$ become $H_2({\bf m}, C_{25},
     C_{35}, C_{45}, C_{55}, 0)$ and $H_2({\bf m}, C_{26},
     C_{36}, C_{46},0, C_{66})$.

\section{Comparison of results for moment equations and simulation}
\subsection{Examples where moment equations and simulation
give good agreement}
Generally agreement between the results for the moment equation
method (MEM) and simulation, in the 
sense that the mean and variance of the voltage as functions of time were reasonably 
close, was obtained when the net mean driving force was large and the variances small.  This is to be expected from the assumptions under
which the moment equations are derived. 

Figure  6 shows one such case where there is excitation only with 
parameter values for synaptic input,  $\bar{g}_e=3$, $\sigma_e=0.0003$, $\tau_e=2$,
$V_E=80$, 50 trials and standard Hodgkin-Huxley parameters as given in the appendix.                                       

For most input parameter sets examined the maximum variance for the MEM was greater than that obtained by simulation. In a very few examples,
particularly for large $\bar{g}_e$ and with a very small time step, the
maximum variance for the simulation solution was greater than that
for the MEM. The approximation for the variance could be
overestimating the true variance, but it could also reflect that either
there is a very small probability that the first spike would be
delayed, advanced or even entirely 
inhibited leading to an increased variance, or that the amplitude in
rare cases is larger during the first spike, also increasing the
variance, but these probabilities are so small that they are not seen
in practice in the simulations.  

\begin{figure}[!h]
\begin{center}
\centerline\leavevmode\epsfig{file=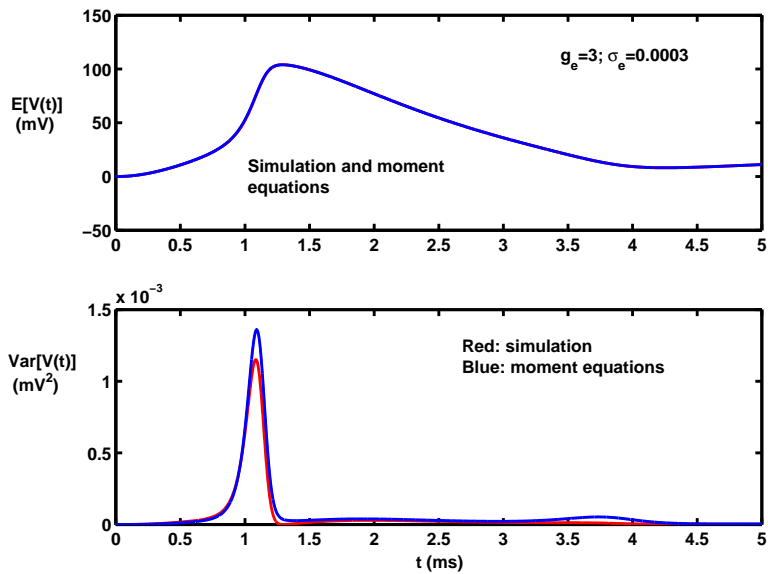,width=4in}
\end{center}
\caption{An example, with excitation only, of the calculated mean and variance of the
voltage $V(t)$ as functions of time of the voltage where there is good agreement between results for simulation (red) and for moment equations (blue).  The input parameters are $\bar{g}_e=3$
and $\sigma_e=0.0003$, with no inhibition and standard HH parameters.  The means for the two methods are indistinguishable.} 
\label{fig:wedge}
\end{figure}

In a second example with reasonable agreement for the two methods
there is synaptic excitation and synaptic inhibition with parameters $\bar{g}_e=3$, $\bar{g}_i=1$, $\sigma_e=0.0003$, $\sigma_i=0.0002$, $\tau_e=2$, $\tau_i=6$,
$V_E=80$, $V_	I=-10$ and standard HH parameters as given in the appendix.                                        
30 simulation runs of 50 trials each were performed.
Again, the mean of $V(t)$ obtained by each method was almost identical, as seen in the top part of Figure 7, where there is one blue curve for the MEM  and 30 red curves for the simulations.

\begin{figure}[!h]
\begin{center}
\centerline\leavevmode\epsfig{file=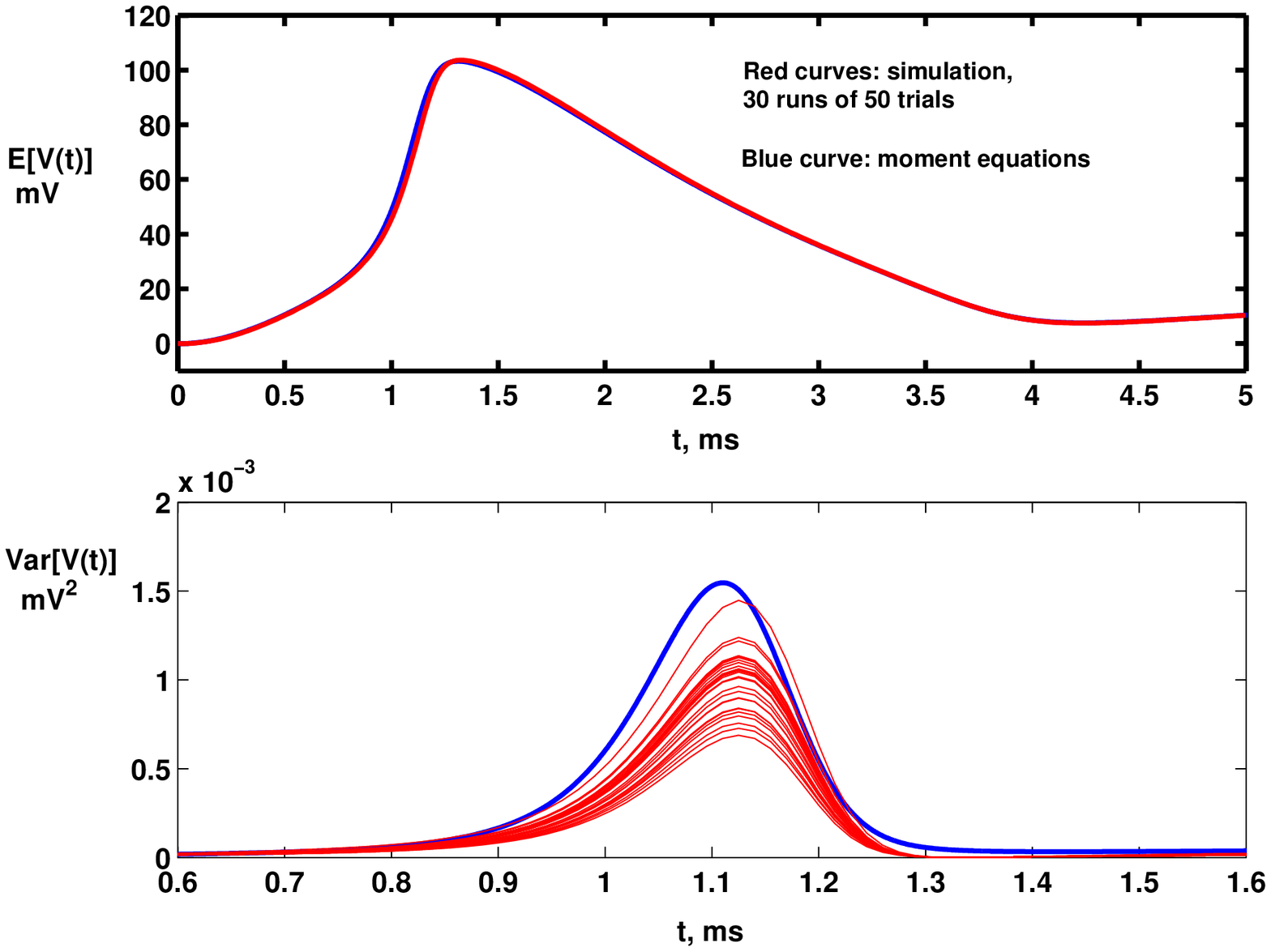,width=4in}
\end{center}
\caption{An example, with both excitation and inhibition, of the calculated mean and variance of the
voltage $V(t)$ as functions of time of the voltage where there is fairly good agreement between results for simulation (red) and for moment equations (blue).  The time scales are different for the mean and the variance.  The principal synaptic input parameters are 
$\bar{g}_e=3$, $\bar{g}_i=1$, $\sigma_e=0.0003$, $\sigma_i=0.0002$.
For remaining parameters see text.  Results are given for 30 runs of 50 trials each. Note the variability in the curves for the variance by simulation (lower figure, red curves). The means for the two methods are practically  indistinguishable.} 
\label{fig:wedge}
\end{figure}

The corresponding results for the variance of $V(t)$ are shown in the 
bottom part of Figure 7. 
For the MEM the maximum variance of $V(t)$
is 0.0015 which is greater than  the maximum variance in each of the
30 simulation runs.  For the latter, the maximum value of all the
maxima is 0.0014, which is 
fairly close to the value for the MEM, and the minimum value is
0.00069, with an average of 0.0010, which is about 33\% less than the
MEM value. For 29 of the 30 runs, simulation values of Var$[V(t)]$
were less than the MEM value for all $t$ and in the remaining case  
the simulation value crossed the MEM value just after the peak value and remained above it for most of the falling phase.

\subsubsection{Occasional spike mode}
The spiking reported in Destexhe et al. (2001) and Fellous et al. (2003)  is classified as  being from
a cell operating in the occasional spike mode (Calvin, 1975), a term
first used with reference to spinal motoneuron spiking. There is insufficient net depolarizing current to give rise to a sustained train of action potentials but occasional large excursions to supra-threshold
states arise due to random synaptic input.

With a small depolarizing current of $\mu=4$, which is less than the critical value for repetitive firing, and with the standard set
of synaptic input parameters reported in column 1 of Table 1 in Destexhe et al. (2001),  
$\bar{g}_e=0.012 \mu$S, $\bar{g}_i=0.057 \mu$S, 
$\sigma_e=0.003$, $\sigma_i=0.0066$ but with
$\tau_e=2$, $\tau_i=6$, and with initially resting 
conditions, a single sample path was generated as depicted in the
top left part of Figure 8. A spike emerged which attained a maximum depolarization
of about  97 mV at approximately 8.5 ms with corresponding sample paths for $g_e$ and $g_i$ shown up to 20 ms in the top right part of Figure 8. Subsequently there are subthreshold  fluctuations, labelled $V_s$ for $t \ge 40$  about a mean of  2.84 mV depicted in the lower left part of the figure. A histogram of voltage values for $V_s$ from $t=40$ to $t=100$ and corresponding histograms of values of the excitatory and inhíbitory conductances are also shown, being similar to those in Figure 2 of Destexhe et al. (2001). 

Starting with the values of all components $V$, $n$, $m$, $h$, $g_e$ and $g_i$ at $t=40$ (beginning of $V_s$), three sets of simulations and
moment equation calculations were performed with the results to 5 ms shown in Figure 9. There is good agreement between the means and variances determined by the two methods.

\begin{figure}[!h]
\begin{center}
\centerline\leavevmode\epsfig{file=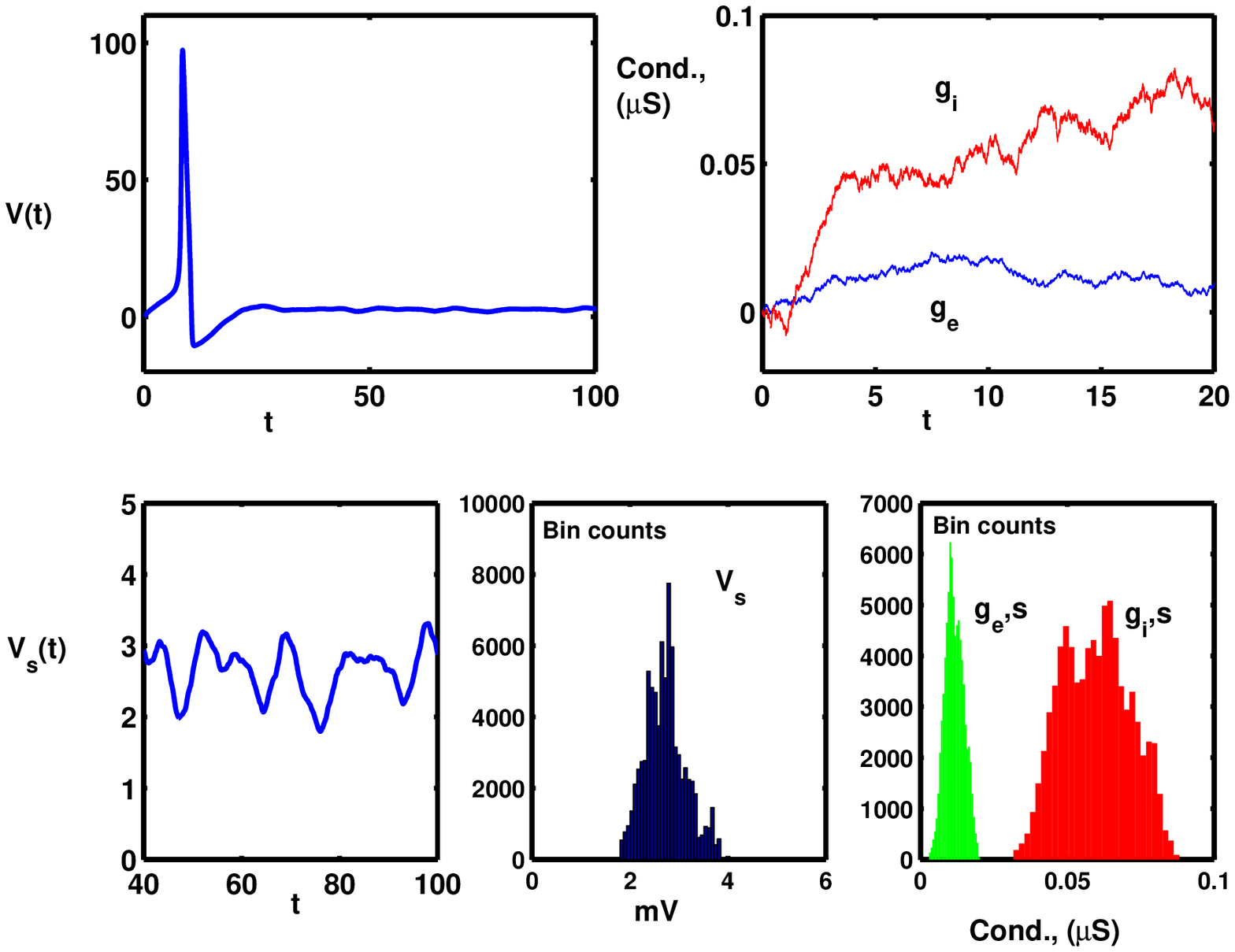,width=5in}
\end{center}
\caption{In the top left figure is shown a spike followed by 
fluctuations in a steady state up to 100 ms. In the top right are shown the 
sample paths of $g_e$ and $g_i$ for the first 20 ms.
After $t=40$ the voltage fluctuations are designated $V_s$ with sample path and histogram of values shown in the first two bottom figures.
In the bottom right figure are the histograms of $g_e$ and $g_i$
during the period from 40 to 100 ms. The latter are comparable to those
in Figures 2 and 3 of Destexhe et al. (2001). For parameter values, see text.} 
\label{fig:wedge}
\end{figure}

\begin{figure}[!h]
\begin{center}
\centerline\leavevmode\epsfig{file=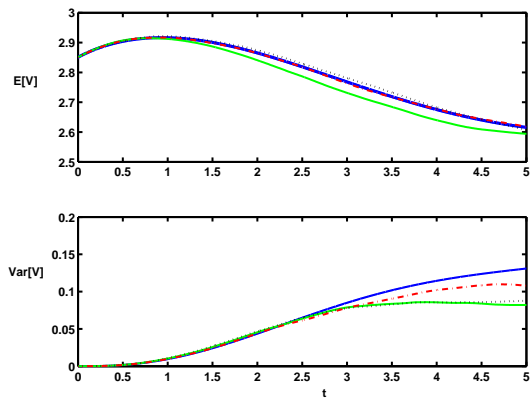,width=2.75in}
\end{center}
\caption{The mean and variance for the subthreshold voltage fluctuations of the previous figure calculated by the MEM (blue curves) and 3 sets of simulations (red, black and green curves).} 
\label{fig:wedge}
\end{figure}

Figure 10 illustrates further the occurrence of the occasional spike mode
over a time period of 400 ms. In all three records the inhibitory steady state conductance is the standard (Destexhe et al. (2001), Table 1, column 1)  value of  $\bar{g}_i=0.057$ and its noise
amplitude is the standard value of  $\sigma_i=0.0066$.
In Figure 10A, there are no spikes with $\mu=4$,  $\bar{g}_e$ at 4 times the standard value of 0.012 and with $\sigma_e=0.003$ which is the standard value.  In the middle record of Figure 10B, there is slightly less additive depolarizing drive with $\mu=3.9$, less mean synaptic excitation 
with $\bar{g}_e$ at 3.6 times the standard value but with considerably
greater noise amplitude of 3.2 times the standard value which is
sufficient to induce occasional spiking. In Figure 10C, there is an
initial singlet spike followed by two doublets with intra-doublet
intervals of about 20 ms. Here the parameters are all as in Figure 10A
(no spikes) but the  
excitatory noise level is 3.25 times the standard value.
With occasional spikes it is clear that the moment method will not
be suitable because the condition of a symmetric distribution of 
component values is not met.

\begin{figure}[!h]
\begin{center}
\centerline\leavevmode\epsfig{file=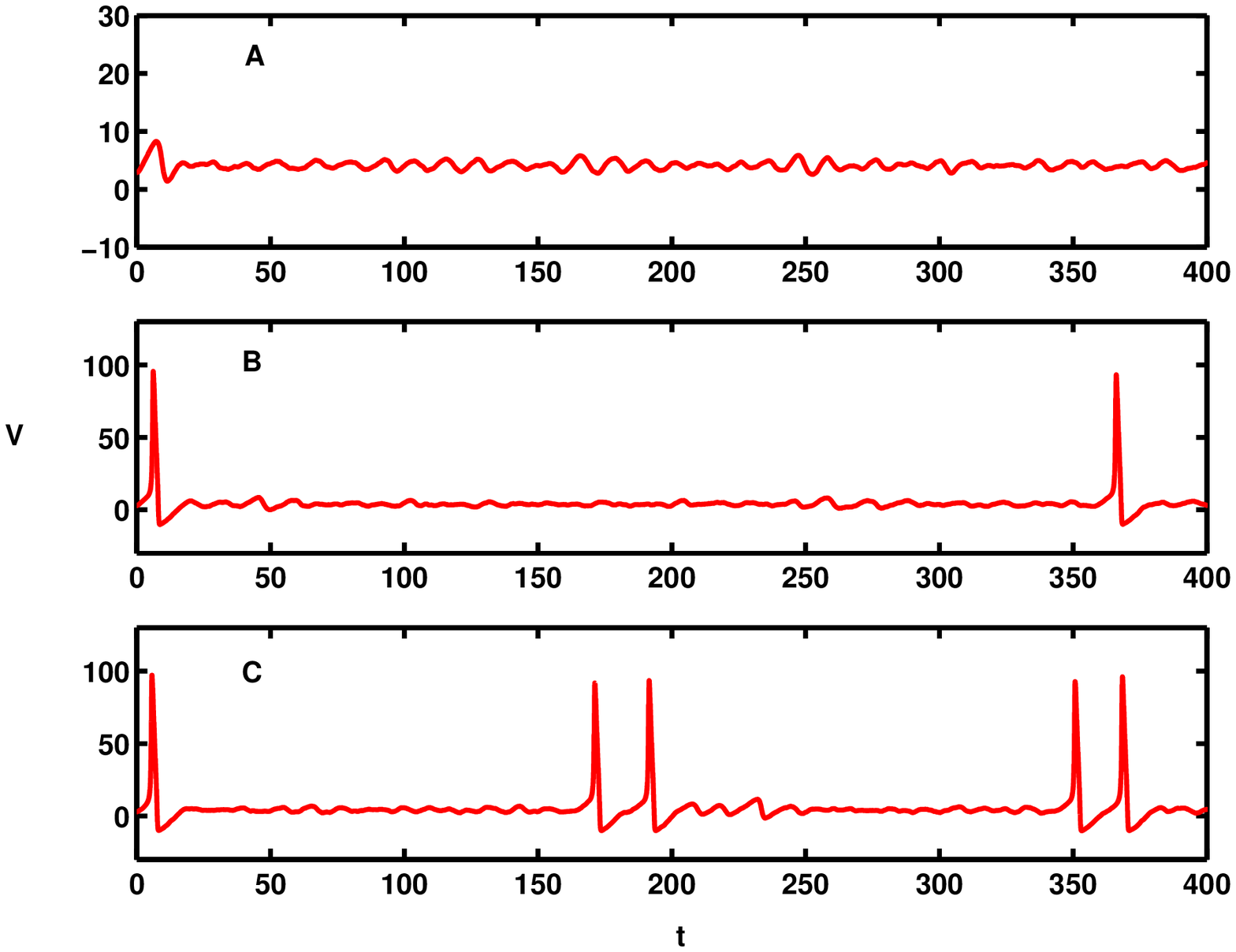,width=2.75in}
\end{center}
\caption{Sample paths of duration 400 ms with added current
of strength $\mu$ and synaptic excitation and inhibition.
{\bf A.}  Only subthreshold fluctuations occur for this period with $\mu=4$, $\bar{g}_e=0.048$ and $\sigma_e=0.003$.  
{\bf B.}  Occasional spike mode in which singlet spikes occur. 
Here $\mu=3.9$,  $\bar{g}_e=0.0432$ and $\sigma_e=0.0096$.  
{\bf C.}  Occasional spike mode in which spikes occur, sometimes in pairs.  Here $\mu=4$,  $\bar{g}_e=0.048$ and $\sigma_e=0.0098$. 
Remaining parameters held fixed - see text.} 
\label{fig:wedge}
\end{figure}

\subsubsection{Input parameters of Destexhe et al. (2001)}
As pointed out in the model description,  Destexhe et al. (2001) 
studied a point HH model augmented with an M-type potassium current  and random excitatory and inhibitory 
synaptic inputs. Apart from the M-type potassium current their
model has the same structure as the one employed here.

It was of interest to see how results for the MEM compared with
those for simulation, despite the lack of knowledge of the complete set of
parameters for the transient sodium and delayed rectifier potassium
current as well as the omission of the M-type current.
We use data in  column 1 of Table 1 of Destexhe et al. (2001) on
properties of a layer 6 pyramidal cell (P-cell) of cat neocortex. A resting
potential given in the text was stated to be -80 mV. 

The P-cell model parameters are as follows.	In Equ. (11) we need   $C$ the whole cell capacitance, which based
on the given membrane area of 34,636 $\mu$m$^2$ is 
0.34636 nF.  If conductances are in $\mu$S, and voltages
are in mV, then currents are in nA. The equilibrium excitatory and
inhibitory conductances for the whole cell are given as
$\bar{g}_e=0.012 \mu$S, $\bar{g}_i=0.057 \mu$S with reversal
potentials of $V_e=80$ mV and $V_i=5$ mV relative to a resting value of $V_R=0$ mV.   The leak conductance, using the value in Destexhe et al. (2001) is $g_L=0.01559 \mu$S, which is about 6 times less than the
value obtained if the standard HH value of 0.3 mS/cm$^2$ is employed.  For the values of 
$\ov{g}_K$ and $\ov{g}_{Na}$, we use the data of Par\'e et al. (1998)
to obtain 3.4636 $\mu$S and 2.4245 $\mu$S, respectively.

Results of a simulation for the P-cell model including synaptic input and with HH  activation and inactivation dynamics and no added
current are shown in Figure 11.  The records have a duration of 100 ms during which the voltage, shown in the top left figure, fluctuates about a mean
of 0.17 mV (above rest) with a standard deviation of 1.6 mV.  The 
distribution of $V$ is  indicated by a histogram in the top right part of Figure 11. The lower two figures show the time courses of the excitatory and inhibitory conductances, $g_e$ and $g_i$.   There are no spikes 
as the fluctuations do not take $V$ to threshold values. The mean in Destexhe et al. (2001) is much higher so that spikes do arise occasionally - called the occasional spike mode as mentioned above.
The values of $g_e$ in Figure 11 are between about 0.001 and   0.023
$\mu$S; those of $g_i$ are between -0.02 $\mu$S  and a maximum of
about 0.09 $\mu$S. That the conductance may become slightly negative
is a minor deficiency of the model equations, being due to the fact
that  the conductances are unrestricted OUPs. (This was allowed for in
Tuckwell et al., 2009). Overall,  the conductance fluctuations are
comparable with those 
in Destexhe et al. (2001), being governed by the same stochastic differential equations.

\begin{figure}[!h]
\begin{center}
\centerline\leavevmode\epsfig{file=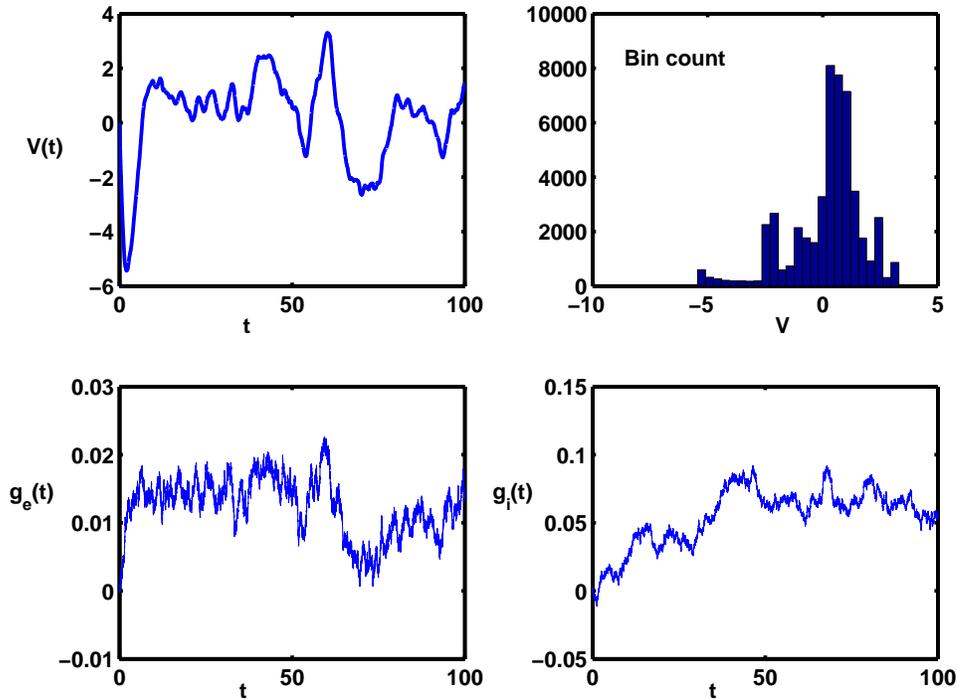,width=5in}
\end{center}
\caption{Properties of one sample path for the P-cell with excitatory and inhibitory synaptic input. The top left figure shows $V$ in mV versus $t$ in ms and the  top right gives a histogram
of the values of $V$ over the 100 ms period.   In the lower two
figures are the corresponding sample paths for the excitatory and
inhibitory conductances. 
For parameters see text.} 
\label{fig:wedge}
\end{figure}

The P-cell model with HH activations and inactivation can be made to
fire by either introducing an additive depolarizing current $\mu$ or by
increasing the ranges of the fluctuations of the synaptic conductances. 
Figure 12 shows two such sets of results where results for simulation compare favorably with those obtained by solving the moment differential
equations.  In both cases there is an  added depolarizing current 
$\mu=10$ $\mu$A/cm$^2$. For Figure 12A, the synaptic steady state
conductances are $\bar{g}_e=1.2$, $\bar{g}_i=5.7$ with corresponding standard deviation parameters  $\sigma_e=0.003$ and $\sigma_i=0.0066$ as above. After a broad small amplitude spike, a steady state 
is attained at a depolarized level with no further spikes. The agreement
of the MEM and simulation for the mean and variance of $V$ is 
very good throughout the spike and immediately afterwards. The variance rises and falls during the rising and falling phases of the spike.
In Figure 12B, the steady state synaptic conductances have the
smaller values $\bar{g}_e=0.12$, $\bar{g}_i=0.57$ with larger corresponding standard deviation parameters  $\sigma_e=0.03$ and $\sigma_i=0.066$. Again a small ampltude broad spike forms
after which a fluctuating steady state is attained. Interesting is the
fact that in this case the variance has two maxima:  it rises on the leading
edge of the spike, decreases around the peak, rises again 
until about half-way down the falling edge and then declines
to a steady state value.  This behavior of the variance is 
confirmed by the calculation by MEM. 
\begin{figure}[!h]
\begin{center}
\centerline\leavevmode\epsfig{file=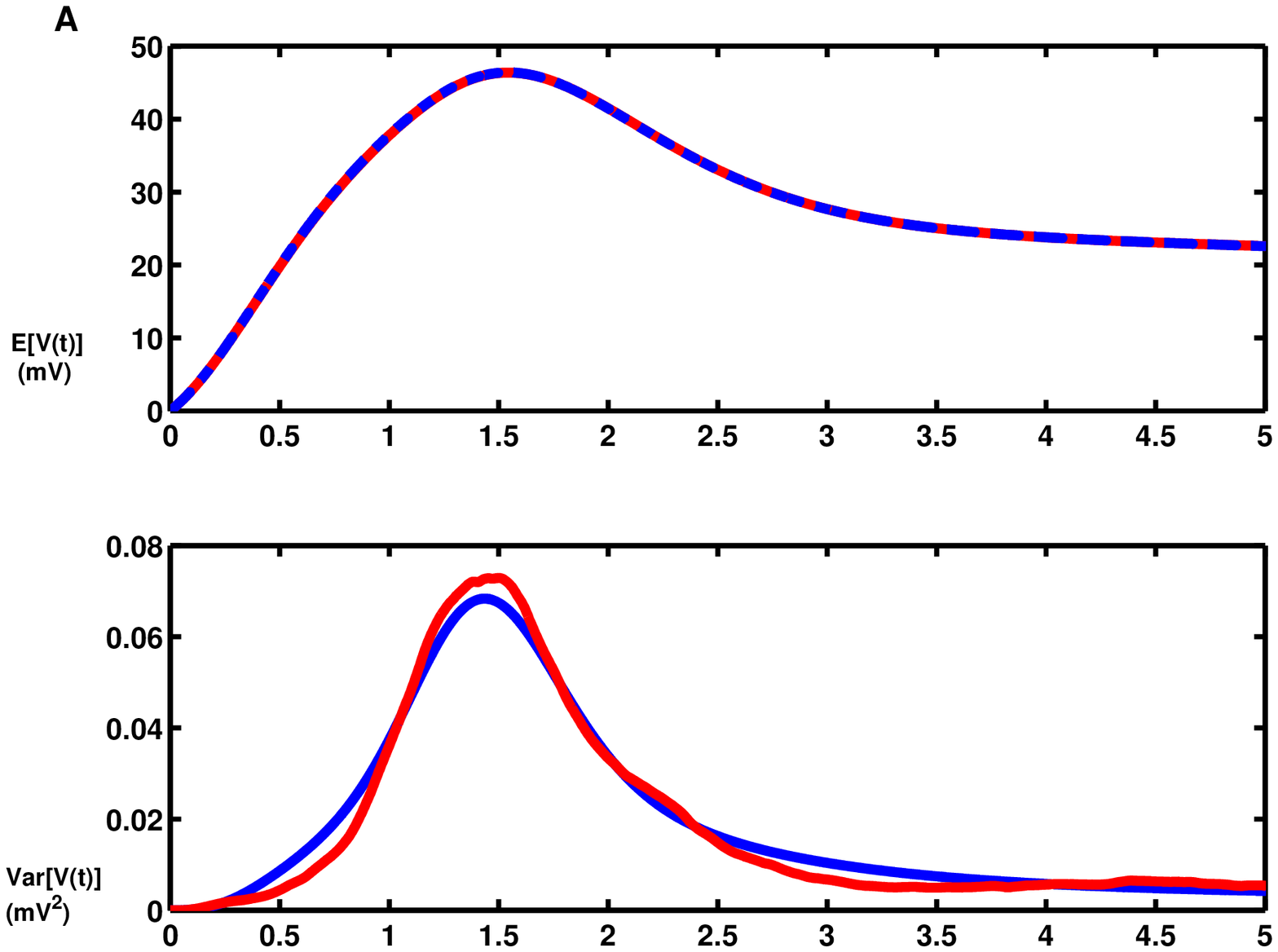,width=2.75in}\epsfig{file=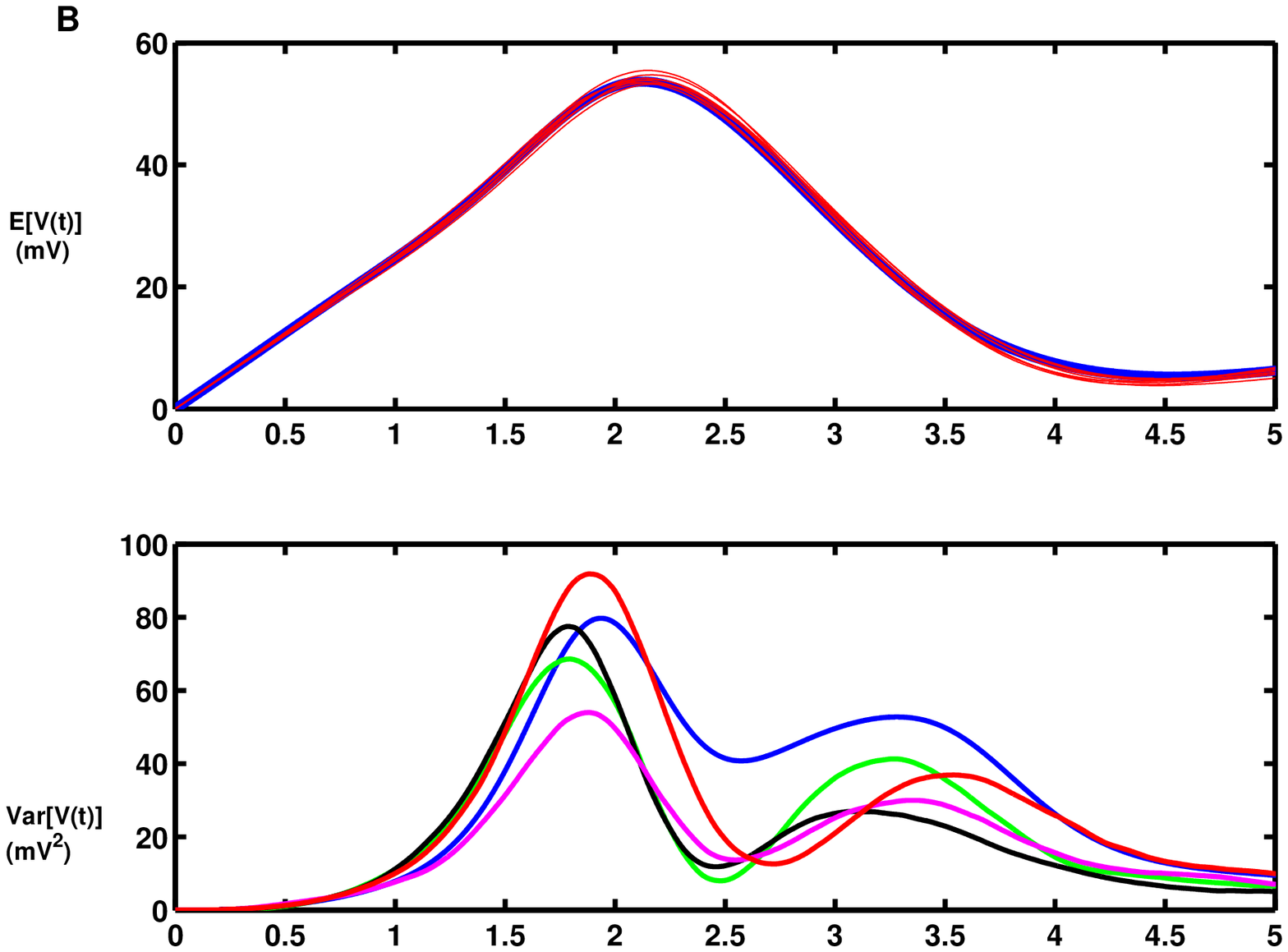,width=2.75in}
\end{center}
\caption{{\bf A}. An example for the P-cell model with an added depolarizing current and strong synaptic excitation and inhibition such that threshold
for spiking is exceeded. Here there is very good agreement between the mean and variance of $V(t)$ determined by simulation (red curves) and by the MEM (blue curves).  {\bf B}.  A second example for the P-cell model with the same depolarizing current as in A but with weaker but noisier synaptic excitation and inhibition. There are 4 sets of simulations with the same number of trials. There is broad agreement between the variance versus time curves from simulation (various colors) and the result from MEM (blue curve). Note the secondary maximum in the variance. For both A and B the means for the two methods are practically indistinguishable. For parameter values see text.} 
\label{fig:wedge}
\end{figure}

 \section{Discussion and conclusions}
The Hodgkin-Huxley (1952) mathematical model for voltage and
current responses in squid axon has formed a cornerstone for
describing the dynamics of neuronal subthreshold and spiking activity
in a large number of types of neurons in many parts of
the nervous system in diverse species. Whereas the original model
involved only three types of ionic current, being leak, potassium and sodium, models of most neurons have been found to require several types of sodium, potassium and calcium currents, each of which has a distinct role in determining a cell's electrophysiological properties. 
The HH system itself is an important starting point in the analysis of
neuronal behavior and is used in the present paper to 
examine the effects of random synaptic activity on repetitive spiking.
Many of the findings are expected to carry over to more complex models.

The present article contains an extension of some previous studies
of the effects of noise in space-clamped HH models. 
Firstly, the results of the inclusion of additive white noise and
excitatory synaptic noise represented by OUPs, on periodic spiking
(Tuckwell et al., 2009).  Secondly, the use of a system  
of deterministic differential equations to find approximately the
first and second-order moments of the HH variables $V, n,
m, h$ 
(Rodriguez and Tuckwell, 2000; Tuckwell and Jost, 2009).  In the first
study it was found that at input signals just greater than critical
values for repetitive (limit cycle) spiking, 
 weak noise could cause a substantial decrease in firing rate and that a
minimum in firing rate occurred as the noise level increased from zero.
In the second type of study, the system of ordinary differential equations
for the approximate moments was solved and the results compared with simulations.

The model employed here  includes synaptic input of the type
supported by experimental analysis and modeling of the voltage response of cat and 
rat neocortical neurons (Destexhe et al, 2001; Fellous et al., 2003). 
The spiking observed in those experiments was occasional (Calvin, 1975) and  the inputs were described  as fast with small amplitude so that they could be represented by continuous random processes (OUPs). We have investigated the effects of these types of
excitatory and inhibitory random inputs on repetitive spiking in the HH model.  However, it is likely that similar qualitative results would
obtain if the random synaptic inputs were purely additive.

The effect of weak random excitatory synaptic input,  investigated
in Section 4.1.2, was  to inhibit spiking induced by a purely excitatory current near the threshold for repetitive firing, confirming results in Section IV of Tuckwell et al. (2009).  Section 4.1.3 contained results for 
 the effects of noisy synaptic input on repetitive spiking induced by the simultaneous application of steady excitatory and steady
inhibitory currents. Here it was demonstrated (Figure 4) that weak noise in
either the excitatory or inhibitory input process or in both could strongly inhibit spiking.  Thus a small amount of noise in any synaptic input channel can cause 
a dimunition of repetitive firing near threshold. Of considerable interest were the findings that with both steady excitatory and inhibitory current,
with no noise in the excitatory input,  increasing the noise only in the inhibitory component from zero, gave rise to a minimum in the firing rate as in the phenomenon of inverse stochastic resonance previously demonstrated for the HH model with additive noise and synaptic excitation only (Figure 5). 

The differential equations for the approximate first and second order
moments of the model were derived in Section 3. The system consists of six equations for the means, six for the variances and fifteen for the covariances giving 27 differential equations in total. These were solved numerically with a Runge-Kutta routine and the results compared with those obtained by simulation.  When the level of excitation is too weak to evoke regular firing, the assumptions for the validity of the moment method are not satisfied and agreement with simulation is generally poor. However, when the firing rate was larger, either with stronger excitatory synaptic input  or with an additional added depolarizing current, and when the noise was of small amplitude, then in a few examples good agreement was found between the two methods over small time intervals, as in the Fitzhugh-Nagumo system with additive noise (Rodriguez and Tuckwell, 1996).

\section*{Acknowledgements}
  We appreciate helpful correspondence with Dr Alain Destexhe at
CNRS, Gif-sur-Yvette, France. The work is part of the Dynamical
Systems Interdisciplinary Network, University of Copenhagen.

\section{Appendix: HH parameters,  coefficients and their derivatives}

\subsection*{Coefficients and their 1st and 2nd derivatives}
Recalling that here $V$ is depolarization and not membrane potential (cf the original  forms in HH(1952)), the coefficients in the
auxiliary equations are the following standard ones. 
\begin{eqnarray*}
\alpha_n(V)= {10-V \o 100[e^{(10-V)/10}-1]}, &&  \beta_n(V)  = {1 \o 8} e^{-V/80},\\
\alpha_m(V) =  {25-V \o 10[e^{(25-V)/10}-1] }, &&   \beta_m(V) = 4e^{-V/18},\\
\alpha_h(V)={7 \o 100} e^{-V/20}, &&  \beta_h(V)=  { 1\o
  e^{(30-V)/10} + 1} . 
\end{eqnarray*}
In the moment equations we require their first  and second order derivatives.  The latter are
\begin{eqnarray*}
\alpha_n'(V)= \frac{ 10-Ve^{  1- \frac{V}{10}}  } {  1000\big(   e^{1- \frac{V}{10}} -1  \big)^2 }, &&
\beta_n'(V)=-\frac{1}{640}{e^{-V/80}}, \\
\alpha_m'(V)=\frac{   10 + e^{\frac{25-V}{10}}(15-V)       }     {100
  (e^{  \frac{25-V}{10}} -1)^2      } ,  && \beta_m'(V)=-\frac{2}{9} e^{-V/18},\\
\alpha_h'(V)=-\frac{7}{2000}e^{-V/20} , &&
 \beta_h'(V)=\frac{   e^{\frac{30-V}{10}  }   }    {10(e^{ \frac{30-V}{10} } + 1)^2}. 
\end{eqnarray*}
	The corresponding second derivatives are found to be
\begin{eqnarray*}
\alpha_n''(V)=  \frac{ 2\beta[(10-\beta V) - 5(\beta e -1)(1-V/10)] }
	                                     { 10^4 (\beta e -1)^3       },&&
\beta_n''(V)=\frac{1}{51200}e^{-V/80} , \\
\alpha_m''(V)= \frac{2\alpha\big(\alpha(15-V) + 10\big) - 10 e^{ \frac{-V}{10}} (25-V)(\alpha-1) }        
	      {  1000(\alpha-1)^3   },&&
\beta_m''(V)= \frac{1}{81}e^{-V/18}, \\
\alpha_h''(V)= \frac{7}{40000} e^{-V/20},&&
\beta_h''(V)= \frac{\gamma}{100}\bigg( \frac{1} { (1+\gamma)^2}    + \gamma \bigg) ,
\end{eqnarray*}
	where
	$$\alpha=e^{\frac{25-V}{10}}, \quad \beta=e^{-\frac{V}{10}} , \quad \gamma=e^{ \frac{30-V}{10}}.  $$
	
	\subsection*{Standard parameters and initial values}
The values of the HH parameters employed in the present model 
are the standard ones:  $C=1$, $\ov{g}_K=36$, $\ov{g}_Na=120$, $g_L=0.3$,
$V_K=-12$, $V_{Na}=115$ and $V_L=10$.  
The initial (resting) values are $V(0)=0$,  $$m(0)=\frac{\alpha_m(0)}
{\alpha_m(0) + \beta_m(0)}, \quad   h(0)=\frac{\alpha_h(0)}
{\alpha_h(0) + \beta_h(0)}, \quad n(0)=\frac{\alpha_n(0)}
{\alpha_n(0) + \beta_n(0)}.$$
The numerical values of the
last three quantities are  0.0529,  0.5961 and 0.3177, respectively.


\section*{References}
\nh Austin, T.D., 2008. The emergence of the deterministic Hodgkin-Huxley equations as a limit from the underlying stochastic ion-channel mechanism. The Annals of Applied Probability 18, 1279-1325.

\nh Bachar, M., Batzel, J.J. and Ditlevsen, S., Eds., 2013. Stochastic Biomathematical Models: with Applications to Neuronal Modeling. 
Lecture Notes in Mathematics 2058. Springer, Berlin. 

\nh  Bashkirtseva, I., Neiman, A. B., Ryashko, L. (2015). Stochastic sensitivity analysis of noise-induced suppression of firing and giant variability of spiking in a Hodgkin-Huxley neuron model. Physical Review E, 91(5), 052920.



\nh Berglund, N., Kuehn, C., 2015. Regularity structures and renormalisation of FitzHugh-Nagumo SPDEs in three space dimensions. arXiv preprint arXiv:1504.02953.

\nh Berg, R.W. and Ditlevsen, S., 2013. Synaptic inhibition and
excitation estimated via the time constant of membrane potential
fluctuations. Journal of Neurophysiology, 110: 1021-1034.

\nh Blair, E.A., Erlanger, J., 1932. Responses of axons to brief shocks. Experimental Biology and Medicine 29, 926-927.

\nh Brink, F., Bronk, D. W., Larrabee, M. G.,1946. Chemical excitation of nerve. Ann. N.Y. A cad. Sci. 47, 457-85.

\nh Burns, B. D., Webb, A. C.,1976. The spontaneous activity of neurones in the cat's
cerebral cortex. Proc. Roy. Soc. Lond. B. 194, 211-23.

\nh Calvin, W.H., 1975. Generation of spike trains in CNS neurons.
Brain Res. 84, 1-22.

\nh Clay, J.R., 2005. Axonal excitability revisited. Progress in Biophysics and Molecular Biology 88, 59-90.

\nh Colquhoun, D., Hawkes, A.G., 1981. On the stochastic properties of single ion channels. Proceedings of the Royal Society of London B: Biological Sciences 211, 205-235.

\nh Deco, G. and Marti, D., 2007. Extended method of moments for deterministic analysis of
stochastic multistable neurodynamical systems. Physical Review 75, 031913.

\nh  Deco, G., Ponce-Alvarez, A., Mantini, D. et al., 2013. Resting-state functional connectivity emerges from structurally and dynamically shaped slow linear fluctuations. J.  Neurosci.  33, 11239-11252.

\nh  Destexhe, A., Rudolph, M. , Fellous, J-M., Sejnowski, T.J., 2001. 
Fluctuating synaptic conductances recreate in vivo-like
activity in neocortical neurons. Neuroscience 107, 13-24.

\nh Ditlevsen, S., Lansky, P., 2005. Estimation of the input parameters in the Ornstein-Uhlenbeck neuronal model. Physical Review E 71,011907.

\nh Ditlevsen, S., 2007. A result on the first-passage time of an Ornstein-Uhlenbeck process. Statistics and Probability Letters 77, 1744-1749.

%

\nh Fellous, J-M., Rudolph, M., Destexhe, A., Sejnowski, T.J., 2003. Synaptic background noise controls the input/
output characteristics of single cells in an in vitro
model of in vivo activity. Neuroscience 122, 811-829.

\nh Finke, C., Vollmer, J., Postnova, S. and Braun, H.A., 2008. Propagation effects of current and conductance noise in a model neuron with subthreshold oscillations. Mathematical Biosciences 214, 109-121.

\nh  Franovi\'c, I., Todorovi\'c, K., Vasovi\'c, N.,Buri\'c, N., 2013. Mean-field approximation of two coupled populations of excitable units. Phys. Rev. E 87, 012922.

\nh Gerstein, G. L., Kiang, N. Y.-S., 1960.  An approach to the quantitative analysis of
electrophysiological data from single neurons. Biophys. J. 1, 15-28.

\nh  Gihman, I. I. and Skorohod, A.V. (1972). {\it Stochastic Differential Equations.} Springer, Berlin.  

\nh Guo, D., 2011. Inhibition of rhythmic spiking by colored noise in neural systems. Cognitive neurodynamics 5, 293-300.

\nh Hasegawa, H., 2009. Population rate codes carried by mean, fluctuation and synchrony of
neuronal firings.  Physica A 388, 499-513. 

\nh Hasegawa, Y., 2015. Variational superposed Gaussian approximation for time-dependent solutions of Langevin equations. Physical Review E 91, 042912.

\nh Hillenbrand, U., 2002. Subthreshold dynamics of the neural membrane potential driven by stochastic synaptic input. Physical Review E, 66, 021909.

\nh  Hodgkin, A.L. and Huxley, A.F. (1952)  A quantitative description of membrane current and its
                     application to conduction and excitation in nerve. {\it J. Physiol. (Lond.)},  {\bf 117} 500-544. 


\nh Horikawa, Y., 1991. Noise effects on spike propagation in the stochastic Hodgkin-Huxley models. Biological cybernetics 66,19-25.

\nh Lecar, H., Nossal, R., 1971. Theory of threshold fluctuations in nerves: I. Relationships between electrical noise and fluctuations in axon firing. Biophysical Journal 11,1048-1067.

\nh Li, Y., Schmid, G., H\"anggi, P., Schimansky-Geier, L., 2010. Spontaneous spiking in an autaptic Hodgkin-Huxley setup. Physical Review E 82, 061907.

\nh  Marreiros, A.C., Kiebel, S.J., Daunizeau, J. et al., 2009. Population dynamics under the Laplace assumption. NeuroImage 44, 701-714.

\nh Monnier, A.M. and Jasper, H.H., 1932. Recherches de la relation entre les potentiels d'action élémentaires et la chronaxie de subordination. Nouvelle démonstration du fonctionnement par “Tout ou Rien” de la fibre nerveuse. C.R. Soc. Biol. 110,, 547-549.


\nh Par\'e, D, Lang, E.J., Destexhe, A., 1998. 
Inhibitory control of somatodendritic
interactions underlying action potentials in
neocortical pyramidal neurons in vivo: an
intracellular and computational study.  Neuroscience 84, 377-402. 

\nh Pecher, C., 1936. Etude statistique des variations spontan\'ees de l'excitabilit\'e d'une fibre nerveuse. C.R. Soc. Biol. 122, 87-91.

\nh Pecher, C., 1937.  Fluctuations ind\'ependantes de l'excitabilit\'e de deux fibres d'une m\^eme nerf. C.R. Soc.
Biol., Paris. 124, 839-842.

\nh Rodriguez, R., Tuckwell, H.C., 1996.  Statistical properties of stochastic nonlinear dynamical models
of single spiking neurons and neural network.  {\it Phys. Rev. E} {\bf 54}, 5585-5589.

\nh Rodriguez, R., Tuckwell, H.C., 1998.
Noisy spiking neurons and networks: useful approximations for
firing probabilities and global behavior. BioSystems 48, 187-194.

\nh Rodriguez, R., Tuckwell, H.C., 2000. 
A dynamical system for the approximate
moments of nonlinear stochastic models
of spiking neurons and networks. 
Mathematical and Computer Modelling 31, 175-180.



\nh Sauer, M., Stannat, W., 2016. Reliability of signal transmission
in stochastic nerve axon equations. Journal of Computational
Neuroscience. To appear. arXiv preprint arXiv:1502.04295.

\nh Sokolowski, T. R., Tka$\check{\rm c}$ik, G., 2015. Optimizing information flow in small genetic networks. IV. Spatial coupling. arXiv preprint arXiv:1501.04015.

\nh Stannat, W., 2013. Stability of travelling waves in stochastic Nagumo equations. arXiv preprint arXiv:1301.6378.

\nh Tuckwell, H.C., 1987. Diffusion approximations to channel noise. Journal of Theoretical Biology 127, 427-438.

\nh Tuckwell, H.C., 2005. Spike trains in a stochastic Hodgkin-Huxley system. BioSystems 80, 25-36.

\nh Tuckwell, H.C., 2008. Analytical and simulation results for the stochastic spatial Fitzhugh-Nagumo model neuron. Neural Computation 20, 3003-3033.

\nh Tuckwell, H.C., Jost, J., 2009.
Moment analysis of the Hodgkin-Huxley system with additive noise.
Physica A 388,  4115-4125.

\nh Tuckwell, H. C., Jost, J. (2012). Analysis of inverse stochastic resonance and the long-term firing of Hodgkin–Huxley neurons with Gaussian white noise. Physica A: Statistical Mechanics and its Applications, 391(22), 5311-5325.

\nh Tuckwell, H.C., Jost, J. and Gutkin, B.S. (2009). 
  Inhibition and modulation of rhythmic neuronal spiking by noise. 
Phys. Rev. E 80, 031907.

\nh  Tuckwell, H.C. and Rodriguez, R. (1998). Analytical and simulation results for stochastic Fitzhugh-Nagumo neurons
and neural networks. {J. Computational Neuroscience} {\bf 5}, 91-113. 

\nh Verveen, A. A., 1960. On the fluctuation of threshold of the nerve fibre. In: Structure and Function of the Cerebral Cortex, pp 282-288.  Eds.,Tower, D.P., Schadé, J.P.). Elsevier,  Amsterdam.

\nh Wenning, G., Hoch, T., Obermayer, K., 2005. Detection of pulses in a colored noise setting. Physical Review E 71, 021902.

\nh White, J.A., Rubinstein, J.T., Kay, A.R., 2000. Channel noise in neurons. Trends in Neurosciences 23, 131-137.

\nh Yu, Y., Shu, Y., McCormick, D.A., 2008.
Cortical action potential backpropagation explains spike
threshold variability and rapid-onset kinetics.
 J. Neurosci. 28., 7260-7272.

\end{document}